\documentclass[aps,preprint,showkeys,showpacs,pre,amsmath,amssymb,amsfonts]{revtex4-1}

\usepackage[dvips]{graphicx}
\usepackage{hyperref}
\usepackage{bm}
\usepackage{color}
\usepackage{latexsym}

\begin{document}

\title{Structural robustness and transport efficiency
of complex networks with degree correlation}

\author{Toshihiro Tanizawa}
\email[E-mail:]{tanizawa@ee.kochi-ct.ac.jp}
\affiliation{Kochi National College of Technology,
200-1 Monobe-Otsu, Nankoku, Kochi 783-8508, Japan}

\begin{abstract} 
 We examine two properties of complex networks, the robustness against
 targeted node removal (attack) and the transport efficiency in terms of
 degree correlation in node connection by numerical evaluation of
 exact analytic expressions.
 We find that, while the assortative correlation enhances
 the structural robustness against attack, the disassortative
 correlation significantly improves the transport efficiency of the
 network under consideration. This finding might shed light on the
 reason why some networks in the real world prefer assortative correlation
 and others prefer disassortative one.
\end{abstract}

\pacs{89.20.Hh,	
      89.75.Fb, 
      89.75.Hc  
}

\keywords{%
complex networks; correlation; robustness; targeted attack; transport
}

\maketitle 

\section{Introduction}
\label{intro}

Many complex systems in the real world can be modeled by complex
networks \cite{Watts:1998vz,Barabasi:1999uu,Albert:2002wu,Newman:2003wd,%
Dorogovtsev:2003Evolut_Networ,Cohen:2005wq,%
Boccaletti:2006gb,Calderelli:2007Large_Scale_Struct,%
newman:2010_networ,cohen:2010_compl_networ,Gao:2011fq%
}.
The cooperative performance of such complex systems relies on the global
connectivity of their components.
These complex systems are placed, however, in an external environment,
where the components or connections could be removed and/or altered.
The analysis of the response of the global connectivity caused
by intentional node removal of the network,
or targeted attacks, has been, therefore,
one of the main issues of the complex network analysis %
\cite{%
Molloy:1995tw,%
albert:2000_error,Cohen:2000vq,Moore:2000ve,%
Callaway:2000vd,Cohen:2001hf,Cohen:2002br,Chung:2002wu,Schwartz:2002dn,Gallos:2005gs,%
Shargel:2003cu,Vazquez:2003fq,%
Paul:2004br,Tanizawa:2005fd,Tanizawa:2006dn,%
Paul:2006cd,Donetti:2006ck,Paul:2007br,buldyrev:2010_catas,Huang:2011cg,%
Dorogovtsev:2003Evolut_Networ,Cohen:2005wq,%
Boccaletti:2006gb,Calderelli:2007Large_Scale_Struct,%
newman:2010_networ,cohen:2010_compl_networ,Gao:2011fq%
}.

Although most of the existing theoretical studies on robustness analysis
of complex networks are based on arguments based only on the degree distribution,
the effects of degree correlation between node connection
begin to be considered recently %
\cite{Serrano:2006ik,Goltsev:2008bf,Shiraki:2010is,Ostilli:2011hb,Tanizawa:2012hh}.
In the early stage of complex network research, however, Newman already
argued that networks in the real world exhibit rather strong tendency, or correlation,
in the connection between nodes of different degrees \cite{Newman:2002jj}.
He introduced the terms, assortative and disassortative correlations,
to describe the tendency of nodes in a network
to make connections between the same degree and
between different degrees, respectively.
Social networks, such as friendship relations, collaboration networks in scientific
research, and so on, tend to be assortative,
and communication networks, such as WWW, neural networks, autonomous system
networks on the Internet, and so on, tend to be disassortative.
Newman also calculated the giant component collapse
for specific kinds of correlated networks against random node removal
and found that assortativity enhances the robustness of networks against
random attack.
Goltsev et al.\ focused on the
evaluation of critical exponents of correlated complex networks
in the vicinity of node percolation transition for the case of
random attack \cite{Goltsev:2008bf}.

Regarding to a robustness analysis against targeted attack,
Schneider et al.\ numerically found that the final robust
network structure have a common ``onion-like'' topology consisting
of a core of highly connected nodes hierarchically
surrounded by rings of nodes with decreasing degree %
\cite{Schneider:2011ip,Herrmann:2011hd}.
In each ring most of the nodes are of the same degree.
Tanizawa et al.\ took an analytic approach to
the robustness analysis against targeted attack \cite{Tanizawa:2012hh}.
Interestingly, they found that the optimal structure against simultaneous random
and targeted attack is very similar
to the ``onion-like'' structure found by Schneider et al.
The optimal structure obtained consists of hierarchically
and weakly interconnected random regular graphs and exhibits
extremely assortative degree correlation.

With respect to the structural robustness against targeted attack,
assortative degree correlation is favorable.
However, there are considerable number of functional
complex networks in the real world that show disassortative degree correlation,
such as WWW, neural networks, autonomous system networks on the
Internet. Within the author's knowledge, there seems to be no plausible
argument for the emergence of disassortative networks.
When we consider that most of the networks that exhibit
disassortativity are communication oriented, in other words, that the
primary function of these networks is information exchange, it is
likely that the efficiency of the information exchange dominates in
determining their network structure.
Gallos et al.\ considered the effects of degree correlation in terms of
fractality in network structure and argued the particle diffusion is
accelerated on the network structure with strong hub repulsion %
\cite{Gallos:2008hd}.

In this paper, we examine the robustness against targeted attack and
the transport efficiency of complex networks in terms of assortative
(disassortative) degree correlation by numerical evaluation of exact
analytic expressions, expecting that the results shed some light on the
reason of the emergence of assortativity (disassortativity) in various
complex networks in the real world.

This paper is organized as follows.
In Sec.\ \ref{theory}, we describe the theoretical framework for
the calculation of the robustness and the transport efficiency of
complex networks with degree correlation under the condition of fixed
degree distribution.
In Sec.\ \ref{results}, we show the results of the calculation
based on the theory described in Sec.\ \ref{theory}.
There we see that the disassortativity enhances the transport efficiency,
while the assortativity enhances the structural robustness against targeted attack.
Section \ref{summary} is for discussion and summary.

\section{Theory}
\label{theory}

\subsection{Degree distribution and degree correlation}

Since we consider the degree correlation in node connection,
we start from the joint degree matrix, $P(k, q)$,
which is the probability that a randomly chosen edge
from a $k$-degree node leads to a $q$-degree node.
For undirected networks,
the symmetry $ P(k, q) = P(q, k) $ holds.
The sum of $ P(k, q) $ over $q$ is the probability
that a randomly chosen edge emanates from a $k$-degree node
and is related to the degree distribution, $P(k)$,
with the normalization condition,
$\sum_k P(k) = 1$,
through the relation,
$ \sum_{q} P(k, q) = k P(k)/ \langle k \rangle$,
where $\langle k \rangle = \sum_k k P(k)$ is the average degree.
By definition, $ \sum_{k,q} P(k, q) = 1$.
If we fix the degree distribution, $P(k)$,
the sum, $\sum_{q} P(k, q)$, has also to be fixed.

The conditional probability, $ P(q|k) $,
that a randomly chosen edge from
a $k$-degree node leads to a $q$-degree node,
which is defined by
$ P(q|k) \equiv P(k, q)/\sum_{q} P(k, q) = P(k, q)/\left( k P(k)/ \langle k \rangle \right)$,
contains the information of the degree correlation of the network under consideration.
This conditional probability, $P(q|k)$, is normalized for each $k$ as
$\sum_{q} P(q|k) = 1$.
For networks without degree correlation, $P(q|k)$ is independent of $k$ and
equals to $q P(q)/ \langle k \rangle$.
This means that the joint degree matrix for uncorrelated networks is separable
and becomes $P(k, q) = \left( k P(k)/\langle k \rangle \right)\left( q P(q)/\langle k \rangle \right)$.

Although the theoretical framework taken in this paper is generic
and applicable to any form of joint degree matrices, $P(k, q)$,
we here consider the cases of scale-free networks where the degree distribution
takes a power-law form, $P(k) \propto k^{-\lambda} \; \left(k = m, m+1, \dots, K-1, K \right)$.
Here we denote the minimum degree as $m$ and the maximum degree as $K$.
In these cases, the joint degree matrix is finite and
the degree dependence of its matrix elements
for uncorrelated (scale-free) networks is represented by
$P_0(k, q) \sim k^{1-\lambda} q^{1-\lambda}$.
By modifying the degree dependence of the joint degree matrix
from this uncorrelated one, we can introduce the degree correlation.
Note that, because of the symmetry of $P(k, q)$ in terms of $k$ and $q$,
we are only allowed to determine the degree dependence of
$P(k, q)$ for the elements corresponding to $q \ge k$.
The matrix elements for $q < k$ are determined from the symmetry and
the normalization condition of $P(k, q)$.

If $P(k, q)$ decreases slower than $P_0(k, q)$ as $q$ increases,
the tendency of nodes to make connection with nodes of larger degree is enhanced.
In this case, nodes are likely to be connected to large degree nodes ($q \gg k$).
Thus the correlation is disassortative.
In contrast, if $P(k, q)$ decreases faster than $P_0(k, q)$ as $q$ increases,
the tendency of nodes to make connection with nodes of larger degrees
becomes smaller.
In this case, nodes are more likely to be connected to nodes of similar degrees ($q \approx k$).
Thus the correlation is assortative.

There are many possibilities to introduce degree correlation.
Here we adopt a simple modification of the joint degree matrix, which is
\begin{equation}
 P(k, q) \sim k^{1-\lambda} q^{1-\lambda + \epsilon} \;\;\;\left( q \ge k \right).
\end{equation}
This modification leads to the conditional probability in the form%
\footnote{Gallos et al.\ took a similar form for the joint degree
probability \cite{Gallos:2008hd}. It should be noted, however, that they
gave the definition for $q \le k$ and the definition of $\epsilon$ is different.}
\begin{equation}
 P(q | k) \sim q^{1-\lambda+\epsilon} \;\;\; \left( q \ge k \right).
\end{equation}
In this modification the degree correlation is controlled by a single parameter $\epsilon$,
which makes analysis sufficiently simple and mathematically tractable.
Since $\epsilon = 0$ means the absence of degree correlation,
nonzero values of $\epsilon$ indicate the existence of degree correlation.
For $\epsilon > 0$, the values of $P(q|k)$ for larger $q$'s increase
from the values of the uncorrelated networks ($\epsilon = 0$), which implies
that the tendency of small degree nodes making connections with large
degree nodes is enhanced. Thus $\epsilon > 0$ represents disassortative
degree correlation.
Similarly, $\epsilon < 0$ represents assortative degree correlation.

\subsection{Giant component collapse due to attack}

In targeted attack,
where the nodes of a network are removed according to the degree of nodes,
the remaining fraction of $k$-degree nodes is reduced
to a factor $b_k \: (0 \le b_k \le 1)$
from the original fraction, $P(k)$.
The total remaining fraction of nodes, $p$,
is calculated as $p = \sum_k b_k P(k) $.

The giant component in a complex network is the largest cluster of connected
nodes, where its normalized size in the network, $S$,
remains finite as the total number of nodes, $N$, becomes infinite.
Non-zero values of $S$ indicate a macroscopic connectivity
of the network under consideration.
The resiliency of $S$ under attack is taken to be a measure of
the robustness of the network.

To calculate the giant component fraction, $S$, under degree correlation,
we take the generating function formalism %
\cite{Callaway:2000vd,Goltsev:2008bf,Tanizawa:2012hh}.
Let $x_k$ be the probability that a randomly chosen link
from a $k$-degree node does not lead to the giant component.
In the limit of $N \to \infty$,
the probability that a node is connected to any node
that has already been connected to other nodes is negligible,
since it is proportional to some negative power of $N$.
Under the condition that the network only consists of trees,
the probabilities, $x_k\; \left( k = m, m+1, \dots, K \right)$,
for non-zero values of $b_k$,
and the node fraction of the giant component, $S$, are determined by
the following set of equations:
\begin{align}
 x_k &= 1 - \sum_{q} b_{q} P(q|k) + \sum_{q} b_{q} P(q|k) \left( x_{q} \right)^{q-1} \label{eq: xk}\\
 S &= p - \sum_k b_k P(k) \left( x_k \right)^k = \sum_k b_k P(k) \left( 1 - (x_k)^k\right).
 \label{eq: S}
\end{align}
Obviously, $x_k = 1$, if all $k$-degree nodes are removed ($b_k = 0$).
The degree correlation is included in the conditional probability,
$P(q|k)$.
Although Eqs.\  (\ref{eq: xk}) and (\ref{eq: S}) contain
$b_k$ and can be applicable to any types of degree based node removal,
we focus on two simple types of attack in this paper:
(i) targeted attack, which is the selective removal of nodes from
the largest degree, and (ii) random attack, which is the uniform removal of nodes
with equal probability.

\subsection{Diffusion on correlated networks}

To investigate the transport property on complex networks with
degree correlation, we consider a simple diffusion process
described by the following simple diffusion equation:%
\footnote{This is the same equation that Gallos et al.\ examined in a
different way \cite{Gallos:2008hd}.}
\begin{equation}
 \frac{d}{d t}\rho_k(t) = - \rho_k(t) + k \sum_{q} P(q|k) \frac{\rho_q(t)}{q},
  \label{eq:diffusion}
\end{equation}
where $\rho_k(t)$ is the density of ``particles'' on a randomly selected
$k$-degree node at time $t$.

The first term in the right hand side of Eq.\ (\ref{eq:diffusion}) represents
the diffusion from the selected $k$-degree node and the second term represents
the total in-coming flow from the $q$-degree neighbor nodes through its $k$ edges.
We assume that the ``particles'' flow out of a $k$-degree node through all of its edges
with equal probability.

By taking vector notation as
$\bm{\rho}(t) = \left\{ \rho_m(t), \rho_{m+1}(t), \dots, \rho_{K-1}(t), \rho_{K}(t) \right\}$
and introducing the conditional probability matrix,
$\mathsf{C} = \left\{ \mathsf{C}_{kq} \right\} = k P(q|k)/q$,
Eq.\ (\ref{eq:diffusion}) can be simply represented as
\begin{equation}
 \frac{d}{d t} \bm{\rho}(t) = -\left( \mathsf{I} - \mathsf{C} \right) \bm{\rho}(t),\label{eq:diffusionEq}
\end{equation}
where $\mathsf{I}$ is the identity matrix.
By the standard diagonalization procedure,
the diffusion equations for each degree are reorganized
to form a set of decoupled independent diffusion equations for each diffusive ``mode''.
The eigenvalues of the matrix $\mathsf{I} - \mathsf{C}$,
which are all non negative in the calculation presented in this paper,
can be interpreted as the diffusion constants for these diffusive modes.
The inverse of the largest positive eigenvalue, which we denote as
$\mu$, is the smallest characteristic time, $T = 1/\mu$, for these diffusive modes
and serves as a good measure for evaluating the efficiency of the diffusion.
Small values of $T$ imply that particles on nodes diffuse in short time intervals, which
we interpret as transport efficiency  of the network under consideration.

\section{Results}
\label{results}

In this section, we present the results of calculation based on the theory
described in the previous section for scale-free networks
with the minimum degree $m = 2$, the maximum degree $K = 20$,
and the exponent in the degree distribution, $\lambda = 1.0$.%
\footnote{
It is well known that $\lambda$ takes a value between 2 and 3
for many scale-free networks in the real world.
Since the main focus of this paper is to investigate the effects of disassortativity,
we take $\lambda = 1.0$ in this paper to make the number of hubs
larger and the effects of disassortativity much more prominent.
We also calculated for the case of scale-free networks
with $m = 2$, $K = 20$, and $\lambda = 2.2$ and obtained similar results.
}

\subsection{Giant component collapse}

The existence of the giant component implies that the global connectivity
in a network is maintained.
As the total remaining node fraction, $p$, decreases,
the node fraction belonging to the giant component, $S$, also decreases.
Vanishing $S$ means that the global connectivity is lost and a network is disintegrated.
It is well known that $S$ vanishes at a certain value of $p$, the threshold,
by percolation phase transition.
The value of the threshold depends on the structure of the network and
the way of node removal.

For targeted attack, the parameters $b_k$ in Eqs.\ (\ref{eq: xk}) and (\ref{eq: S}) are taken to be
\begin{equation}
 b_k  =
  \begin{cases}
   1 & (m \le k < \tilde{K}), \\
   b & (k = \tilde{K}), \\
   0 & (\tilde{K} < k \le K),
  \end{cases}
\end{equation}
where $\tilde{K}$ is the maximum degree of partially remaining nodes.
The values of $\tilde{K}$ and $b$ are determined by the total remaining node
fraction through $p = \sum_{k = m}^{\tilde{K}} b_k P(k)$.
For random attack, the node removal is uniform and all $b_k$'s are equal
to the total remaining node fraction, $p$.

 \begin{figure}
  \begin{center}
   \includegraphics[clip,width=15cm]{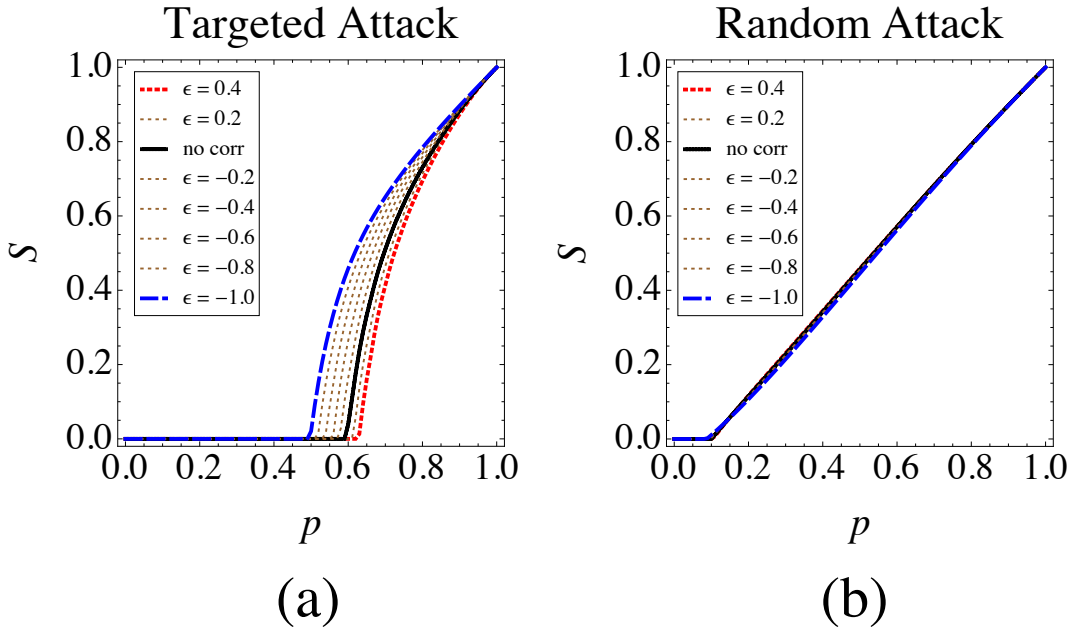}
  \end{center}
  \caption{(Color online) Plots of the giant component fraction, $S$, as a function of remaining node
  fraction, $p$, for scale-free networks with $m = 2$, $K = 20$, and $\lambda = 1.0$
  that have different values of $\epsilon$ from $\epsilon = -1.0$ (assortative, in blue) to
  $\epsilon = 0.4$ (disassortative, in red) through $\epsilon = 0$ (no correlation, in black).
  The plot (a) shows the giant component collapse due to targeted attack
  and the plot (b) shows the giant component collapse due to random attack.
  Note that, while the degree correlation plays no role in the case of random attack,
  the assortative (disassortative) degree correlation enhances (reduces) the resiliency
  of the giant component.}
  \label{fig:GC}
 \end{figure}

In Fig.\ \ref{fig:GC}, we plot the giant component fraction $S$
of scale-free networks with $m = 2$, $K = 20$, and $\lambda = 1.0$
as a function of the remaining node fraction, $p$,
for various values of $\epsilon$ from $\epsilon = -1.0$ (assortative) to
$\epsilon = 0.4$ (disassortative) through $\epsilon = 0$ (no correlation)
against targeted attack and random attack.
We see from Fig.\ \ref{fig:GC} (b) that
the degree correlation plays almost no role in random attack.
In contrast, the degree correlation significantly affects the decrease of $S$
against targeted attack (Fig.\ \ref{fig:GC} (a)).
As the assortativity becomes stronger ($\epsilon < 0$),
the steep drop in $S$ takes place at smaller values of $p$.
Introducing assortative degree correlation thus makes a network more resilient against targeted attack.
    
  \begin{figure}
   \begin{center}
    \includegraphics[clip,width=9cm]{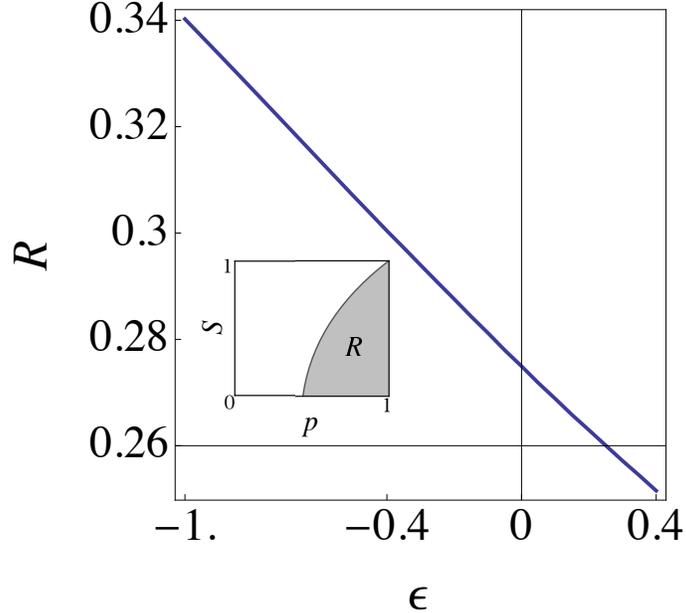}
   \end{center}
   \caption{(Color online) Plot of the robustness measure, $R$, defined in the inset of the plot and
   calculated from the plot, Fig.\ \ref{fig:GC} (a) as a function of $\epsilon$.
   As the value of $\epsilon$ increases from $-1.0$ (assortative) to $0.4$ (disassortative),
   the robustness measure, $R$, decreases almost linearly, which implies the scale-free
   network becomes more vulnerable against targeted attack as the degree correlation tends
   to disassortative.}
   \label{fig:r measure}
  \end{figure}

As discussed in \cite{Schneider:2011ip,Herrmann:2011hd,Tanizawa:2012hh},
the giant component begins to collapse faster for random attack as the assortativity sets in,
while the remaining node threshold decreases slightly with the increase of assortativity.
Therefore, the remaining node threshold does not serve as a good measure for the robustness.
To compare the robustness against attack in terms of degree correlation,
we take the area under the decreasing curve of $S$
as the robustness measure, which represents the resiliency of the giant component due to
node removal much more properly.%
\footnote{This measure is the same as those used in
\cite{Schneider:2011ip,Herrmann:2011hd,Tanizawa:2012hh}}
(See the inset of Fig.\ \ref{fig:r measure}.)
We plot the robustness measure, $R$, calculated
from the curves of $S$ against targeted attack as a function of $\epsilon$
in Fig.\ \ref{fig:r measure}.
As the value of $\epsilon$ increases from $-1.0$ (assortative) to $0.4$ (disassortative),
the robustness measure, $R$, decreases almost linearly, which implies the scale-free
network becomes more vulnerable against targeted attack as the degree correlation tends
to disassortative.
If the robustness against targeted attack is the first priority
in designing network structure, the degree correlation should be assortative.

\subsection{Transport efficiency}

Networks in the real world are not static objects.
There are flows of material, energy, information on the networks.
The transport efficiency of these flows,
as well as the structural robustness,
should also be one of the crucial functionalities of complex networks.

To draw an overall picture of transport efficiency with respect to
degree correlation, we examine the simple diffusion equation,
Eq.\ (\ref{eq:diffusion}), described in the previous section.

  \begin{figure}
   \begin{center}
    \includegraphics[clip,width=15cm]{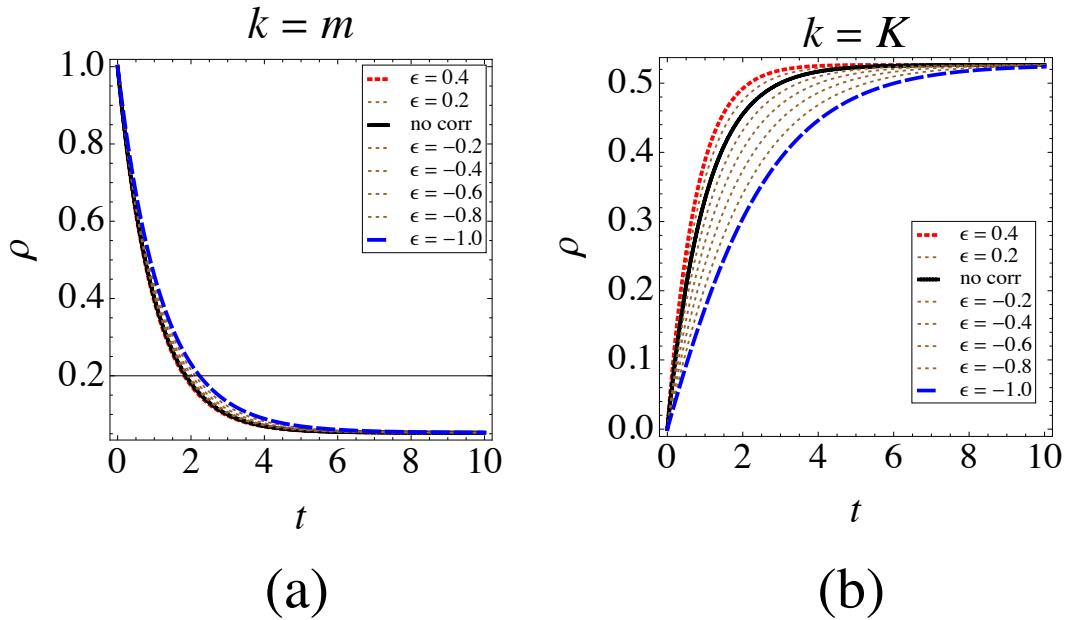}
   \end{center}
   \caption{(Color online) Plots of the time dependence of the density of diffusing particles
   on (a) a node of the minimum degree, $m$, and (b) a node of the maximum degree,
   $K$, in scale-free networks with $m = 2$, $K = 20$, and $\lambda = 1.0$
   according to the diffusion equation, Eq.\ (\ref{eq:diffusionEq}), under
   the initial condition that all particles are uniformly distributed on nodes
   of the minimum degree.
   Thus the particles diffuse from the smallest degree nodes to all the other
   larger degree nodes up to the maximum degree.
   Note that particles diffuse faster as the correlation parameter $\epsilon$
   increases from $-1.0$ (assortative, in blue) to $0.4$ (disassortative, in red).}
   \label{fig:diffusion mK}
  \end{figure}

In Fig.\ \ref{fig:diffusion mK}, we plot the densities of diffusing particles
on a node of the minimum degree and a node of the maximum degree
in scale-free networks with $m = 2$, $K = 20$, and $\lambda = 1.0$ as
a function of time, $t$.
In this calculation, all the particles are uniformly
distributed on the nodes of minimum degree. Thus the particles flow out
of minimum degree nodes and diffuse to other nodes of larger degree.
Figure \ref{fig:diffusion mK} (b) shows the increase of the density on
a node of maximum degree. From this plot, we can see that the density
reaches its maximum faster as the degree correlation becomes disassortative.
This is because, in disassortative networks,
the tendency of minimum degree nodes making connection
to the maximum degree nodes is larger and
that the particles on minimum degree nodes can easily diffuse
to larger degree nodes through these connections.

  \begin{figure}
   \begin{center}
    \includegraphics[clip,width=15cm]{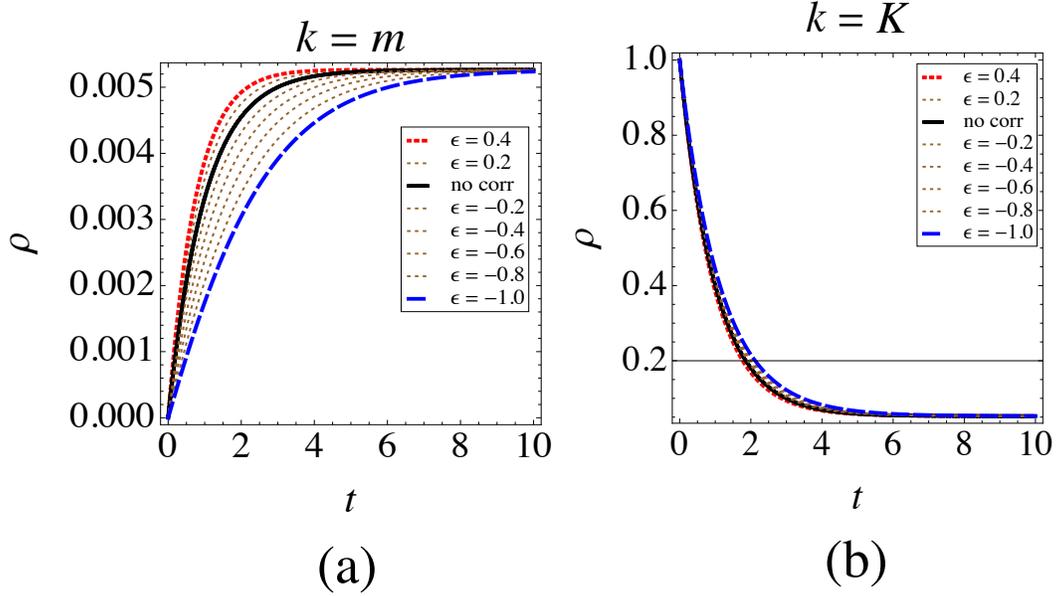}
   \end{center}
   \caption{(Color online) Plots of the time dependence of the density of diffusing particles
   on (a) a node of the minimum degree, $m$, and (b) a node of the maximum degree,
   $K$, in scale-free networks with $m = 2$, $K = 20$, and $\lambda = 1.0$
   according to the diffusion equation, Eq.\ (\ref{eq:diffusionEq}), under
   the initial condition that all particles are uniformly distributed on nodes
   of the maximum degree.
   Thus the particles diffuse from the largest degree nodes to all the other
   smaller degree nodes down to the minimum degree.
   Note that particles diffuse faster as the correlation parameter $\epsilon$
   increases from $-1.0$ (assortative, in blue) to $0.4$ (disassortative, in red).}
   \label{fig:diffusion Km}
  \end{figure}

In Fig.\ \ref{fig:diffusion Km}, we plot the same diffusion process except for
the initial condition, in which all particles are uniformly
distributed on nodes of maximum degree.
Also in this case, the particles diffuse faster in disassortative networks.%
\footnote{
The density, $\rho_k(t)$, in the diffusion equation, Eq.\ (\ref{eq:diffusion}),
is the density of ``particles'' on each node.
Since the number of nodes of $m$ degree is much larger than the number of nodes
of $K$ degree in scale-free networks, the final values of $\rho_m$ plotted in
Fig.\ \ref{fig:diffusion Km} (a) are much smaller than the final values of $\rho_K$
plotted in Fig.\ \ref{fig:diffusion mK} (b).
}

  \begin{figure}
   \begin{center}
    \includegraphics[clip,width=7cm]{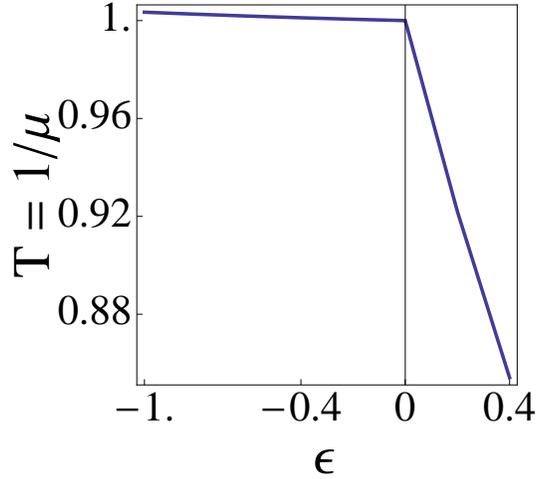}
   \end{center}
   \caption{(Color online) Plot of the characteristic time for diffusion, $T$, defined by the inverse
   of the largest positive eigenvalue, $\mu$, of the coefficient matrix,
   $\mathsf{I} - \mathsf{C}$.
   As the value of $\epsilon$ increases from $-1.0$ (assortative) to $0.4$ (disassortative),
   the characteristic time, $T$, decreases almost linearly with a bend at
   $\left(\epsilon, T \right) = \left( 0, 1 \right)$ (no correlation).
   The decreasing rate in $T$ is much steeper in the disassortative region ($\epsilon > 0$)
   than that in the assortative region ($\epsilon < 0$).}
   \label{fig:charact time}
  \end{figure}

The inverse of the largest positive eigenvalues, $\mu$, of the coefficient matrix,
$\mathsf{I} - \mathsf{C}$, of the diffusion equation, Eq. (\ref{eq:diffusionEq}),
can be taken as a measure the characteristic diffusion time, $T (= 1/\mu)$,
the behavior of which we plot in Fig.\ \ref{fig:charact time} as a function of
$\epsilon$.
It is interesting that the decrease in $T$ has a bend at
$\left(\epsilon, T \right) = \left( 0, 1 \right)$ (no correlation) and
gets steeper in disassortative region ($\epsilon > 0$).
This means that the diffusion takes place much faster in disassortative networks.
Thus the transport efficiency favors disassortative degree correlation.

\section{Discussion and summary}
\label{summary}

Structural robustness and transport efficiency are both important
qualities for the proper functioning of a complex network in the real world.
From the results presented in this paper,
we see that assortative correlation is favorable for structural robustness,
whereas disassortative correlation is favorable for transport efficiency.

As Newman pointed out,
both assortative and disassortative scale-free networks
occur in the real world\cite{Newman:2002jj}.
Social networks, such as friendship relations, collaboration networks in scientific
research, and so on, tend to be assortative,
and communication networks, such as WWW, neural networks, autonomous system
networks on the Internet, and so on, tend to be disassortative.

Within the author's knowledge, there is no plausible argument for
the emergence of this difference.
According to the results obtained here,
one possible argument might be the following.
If the required time scale of transport on a network for proper
functioning is not so fast and the structural robustness against
several types of attack is the first priority,
the resulting network favors assortative degree correlation.
In contrast, the transport efficiency is crucial and mandatory,
the resulting network favors disassortative degree correlation.
The existing examples we have at present seem to fit this reasoning.

The results presented in this paper is, however, obtained from
an application of the theory to very limited cases.
To confirm the validity of the argument about the emergence of
degree correlation, we need more extensive and thorough investigation,
which will be open to our future work.

In summary, we examine the effect of degree correlation
on structural robustness and transport efficiency of scale-free networks
by numerical calculation based on exact analytic equations.
The transport efficiency is significantly enhanced
when the degree correlation becomes disassortative,
whereas the structural robustness against targeted attack is improved
as the assortative degree correlation becomes stronger.
These opposite roles of degree correlation played in
structural robustness and transport efficiency possibly provide
a plausible argument for the emergence of various degree correlation
in complex networks in the real world.

\acknowledgments
The author is grateful to Prof.\ Small and Prof.\ Nakamura for
suggesting him to submit his work to NOLTA.
He also thanks to the financial support of the Grant-in-Aid for
Scientific Research (C) (No. 24540419). 


\begin{thebibliography}{45}%
\makeatletter
\providecommand \@ifxundefined [1]{%
 \@ifx{#1\undefined}
}%
\providecommand \@ifnum [1]{%
 \ifnum #1\expandafter \@firstoftwo
 \else \expandafter \@secondoftwo
 \fi
}%
\providecommand \@ifx [1]{%
 \ifx #1\expandafter \@firstoftwo
 \else \expandafter \@secondoftwo
 \fi
}%
\providecommand \natexlab [1]{#1}%
\providecommand \enquote  [1]{``#1''}%
\providecommand \bibnamefont  [1]{#1}%
\providecommand \bibfnamefont [1]{#1}%
\providecommand \citenamefont [1]{#1}%
\providecommand \href@noop [0]{\@secondoftwo}%
\providecommand \href [0]{\begingroup \@sanitize@url \@href}%
\providecommand \@href[1]{\@@startlink{#1}\@@href}%
\providecommand \@@href[1]{\endgroup#1\@@endlink}%
\providecommand \@sanitize@url [0]{\catcode `\\12\catcode `\$12\catcode
  `\&12\catcode `\#12\catcode `\^12\catcode `\_12\catcode `\%12\relax}%
\providecommand \@@startlink[1]{}%
\providecommand \@@endlink[0]{}%
\providecommand \url  [0]{\begingroup\@sanitize@url \@url }%
\providecommand \@url [1]{\endgroup\@href {#1}{\urlprefix }}%
\providecommand \urlprefix  [0]{URL }%
\providecommand \Eprint [0]{\href }%
\@ifxundefined \urlstyle {%
  \providecommand \doi  [0]{\begingroup \@sanitize@url \@doi}%
  \providecommand \@doi [1]{\endgroup \@@startlink {\doibase
  #1}doi:\discretionary {}{}{}#1\@@endlink }%
}{%
  \providecommand \doi  [0]{doi:\discretionary{}{}{}\begingroup
  \urlstyle{rm}\Url }%
}%
\providecommand \doibase [0]{http://dx.doi.org/}%
\providecommand \Doi [0]{\begingroup \@sanitize@url \@Doi }%
\providecommand \@Doi  [1]{\endgroup\@@startlink{\doibase#1}\@@Doi}%
\providecommand \@@Doi [1]{#1\@@endlink}%
\providecommand \selectlanguage [0]{\@gobble}%
\providecommand \bibinfo  [0]{\@secondoftwo}%
\providecommand \bibfield  [0]{\@secondoftwo}%
\providecommand \translation [1]{[#1]}%
\providecommand \BibitemOpen [0]{}%
\providecommand \bibitemStop [0]{}%
\providecommand \bibitemNoStop [0]{.\EOS\space}%
\providecommand \EOS [0]{\spacefactor3000\relax}%
\providecommand \BibitemShut  [1]{\csname bibitem#1\endcsname}%
\bibitem [{\citenamefont {Watts}\ and\ \citenamefont
  {Strogatz}(1998)}]{Watts:1998vz}%
  \BibitemOpen
  \bibfield  {author} {\bibinfo {author} {\bibfnamefont {D.~J.}\ \bibnamefont
  {Watts}}\ and\ \bibinfo {author} {\bibfnamefont {S.~H.}\ \bibnamefont
  {Strogatz}},\ }\href@noop {} {\bibfield  {journal} {\bibinfo  {journal}
  {Nature (London)},\ }\textbf {\bibinfo {volume} {393}},\ \bibinfo {pages}
  {440} (\bibinfo {year} {1998})}\BibitemShut {NoStop}%
\bibitem [{\citenamefont {Barab{\'a}si}\ and\ \citenamefont
  {Albert}(1999)}]{Barabasi:1999uu}%
  \BibitemOpen
  \bibfield  {author} {\bibinfo {author} {\bibfnamefont {A.-L.}\ \bibnamefont
  {Barab{\'a}si}}\ and\ \bibinfo {author} {\bibfnamefont {R.}~\bibnamefont
  {Albert}},\ }\href@noop {} {\bibfield  {journal} {\bibinfo  {journal}
  {Science},\ }\textbf {\bibinfo {volume} {286}},\ \bibinfo {pages} {509}
  (\bibinfo {year} {1999})}\BibitemShut {NoStop}%
\bibitem [{\citenamefont {Albert}\ and\ \citenamefont
  {Barab{\'a}si}(2002)}]{Albert:2002wu}%
  \BibitemOpen
  \bibfield  {author} {\bibinfo {author} {\bibfnamefont {R.}~\bibnamefont
  {Albert}}\ and\ \bibinfo {author} {\bibfnamefont {A.-L.}\ \bibnamefont
  {Barab{\'a}si}},\ }\href@noop {} {\bibfield  {journal} {\bibinfo  {journal}
  {Rev. Mod. Phys.},\ }\textbf {\bibinfo {volume} {74}},\ \bibinfo {pages} {47}
  (\bibinfo {year} {2002})}\BibitemShut {NoStop}%
\bibitem [{\citenamefont {Newman}(2003)}]{Newman:2003wd}%
  \BibitemOpen
  \bibfield  {author} {\bibinfo {author} {\bibfnamefont {M.~E.~J.}\
  \bibnamefont {Newman}},\ }\href@noop {} {\bibfield  {journal} {\bibinfo
  {journal} {SIAM Review},\ }\textbf {\bibinfo {volume} {45}},\ \bibinfo
  {pages} {167} (\bibinfo {year} {2003})}\BibitemShut {NoStop}%
\bibitem [{\citenamefont {Dorogovtsev}\ and\ \citenamefont
  {Mendes}(2003)}]{Dorogovtsev:2003Evolut_Networ}%
  \BibitemOpen
  \bibfield  {author} {\bibinfo {author} {\bibfnamefont {S.~N.}\ \bibnamefont
  {Dorogovtsev}}\ and\ \bibinfo {author} {\bibfnamefont {J.~F.~F.}\
  \bibnamefont {Mendes}},\ }\href@noop {} {\emph {\bibinfo {title} {Evolution
  of Networks: From Biological Nets to the Internet and WWW}}}\ (\bibinfo
  {publisher} {Oxford University Press},\ \bibinfo {address} {New York},\
  \bibinfo {year} {2003})\BibitemShut {NoStop}%
\bibitem [{\citenamefont {Cohen}\ \emph {et~al.}(2005)\citenamefont {Cohen},
  \citenamefont {Havlin},\ and\ \citenamefont {Ben-Avraham}}]{Cohen:2005wq}%
  \BibitemOpen
  \bibfield  {author} {\bibinfo {author} {\bibfnamefont {R.}~\bibnamefont
  {Cohen}}, \bibinfo {author} {\bibfnamefont {S.}~\bibnamefont {Havlin}}, \
  and\ \bibinfo {author} {\bibfnamefont {D.}~\bibnamefont {Ben-Avraham}},\
  }\href@noop {} {\emph {\bibinfo {title} {{Structural Properties of Scale-Free
  Networks, in Handbook of Graphs and Networks: From the Genome to the Internet
  (Chapter 4)}}}}\ (\bibinfo  {publisher} {Wiley-VCH Verlag GmbH \& Co.\
  KGaA},\ \bibinfo {address} {Weinheim, FRG},\ \bibinfo {year}
  {2005})\BibitemShut {NoStop}%
\bibitem [{\citenamefont {Boccaletti}\ \emph {et~al.}(2006)\citenamefont
  {Boccaletti}, \citenamefont {Latora}, \citenamefont {Moreno}, \citenamefont
  {Chavez},\ and\ \citenamefont {Hwang}}]{Boccaletti:2006gb}%
  \BibitemOpen
  \bibfield  {author} {\bibinfo {author} {\bibfnamefont {S.}~\bibnamefont
  {Boccaletti}}, \bibinfo {author} {\bibfnamefont {V.}~\bibnamefont {Latora}},
  \bibinfo {author} {\bibfnamefont {Y.}~\bibnamefont {Moreno}}, \bibinfo
  {author} {\bibfnamefont {M.}~\bibnamefont {Chavez}}, \ and\ \bibinfo {author}
  {\bibfnamefont {D.-U.}\ \bibnamefont {Hwang}},\ }\href@noop {} {\bibfield
  {journal} {\bibinfo  {journal} {Physics Reports},\ }\textbf {\bibinfo
  {volume} {424}},\ \bibinfo {pages} {175} (\bibinfo {year}
  {2006})}\BibitemShut {NoStop}%
\bibitem [{\citenamefont {Calderelli}\ and\ \citenamefont
  {Vespignani}(2007)}]{Calderelli:2007Large_Scale_Struct}%
  \BibitemOpen
  \bibfield  {author} {\bibinfo {author} {\bibfnamefont {G.}~\bibnamefont
  {Calderelli}}\ and\ \bibinfo {author} {\bibfnamefont {A.}~\bibnamefont
  {Vespignani}},\ }\href@noop {} {\emph {\bibinfo {title} {Large Scale
  Structure and Dynamics of Complex Networks}}}\ (\bibinfo  {publisher} {World
  Scientific},\ \bibinfo {address} {Singapore},\ \bibinfo {year}
  {2007})\BibitemShut {NoStop}%
\bibitem [{\citenamefont {Newman}(2010)}]{newman:2010_networ}%
  \BibitemOpen
  \bibfield  {author} {\bibinfo {author} {\bibfnamefont {M.~E.~J.}\
  \bibnamefont {Newman}},\ }\href@noop {} {\emph {\bibinfo {title} {Networks:
  An Introduction}}}\ (\bibinfo  {publisher} {Oxford University Press},\
  \bibinfo {address} {New York},\ \bibinfo {year} {2010})\BibitemShut {NoStop}%
\bibitem [{\citenamefont {Cohen}\ and\ \citenamefont
  {Havlin}(2010)}]{cohen:2010_compl_networ}%
  \BibitemOpen
  \bibfield  {author} {\bibinfo {author} {\bibfnamefont {R.}~\bibnamefont
  {Cohen}}\ and\ \bibinfo {author} {\bibfnamefont {S.}~\bibnamefont {Havlin}},\
  }\href@noop {} {\emph {\bibinfo {title} {Complex Networks: Structure,
  Robustness and Function}}}\ (\bibinfo  {publisher} {Cambridge University
  Press},\ \bibinfo {address} {Cambridge, UK},\ \bibinfo {year}
  {2010})\BibitemShut {NoStop}%
\bibitem [{\citenamefont {Gao}\ \emph {et~al.}(2011)\citenamefont {Gao},
  \citenamefont {Buldyrev}, \citenamefont {Havlin},\ and\ \citenamefont
  {Stanley}}]{Gao:2011fq}%
  \BibitemOpen
  \bibfield  {author} {\bibinfo {author} {\bibfnamefont {J.}~\bibnamefont
  {Gao}}, \bibinfo {author} {\bibfnamefont {S.}~\bibnamefont {Buldyrev}},
  \bibinfo {author} {\bibfnamefont {S.}~\bibnamefont {Havlin}}, \ and\ \bibinfo
  {author} {\bibfnamefont {H.~E.}\ \bibnamefont {Stanley}},\ }\href@noop {}
  {\bibfield  {journal} {\bibinfo  {journal} {Phys. Rev. Lett.},\ }\textbf
  {\bibinfo {volume} {107}},\ \bibinfo {pages} {195701} (\bibinfo {year}
  {2011})}\BibitemShut {NoStop}%
\bibitem [{\citenamefont {Molloy}\ and\ \citenamefont
  {Reed}(1995)}]{Molloy:1995tw}%
  \BibitemOpen
  \bibfield  {author} {\bibinfo {author} {\bibfnamefont {M.}~\bibnamefont
  {Molloy}}\ and\ \bibinfo {author} {\bibfnamefont {B.}~\bibnamefont {Reed}},\
  }\href@noop {} {\bibfield  {journal} {\bibinfo  {journal} {Random Structures
  {\&} Algorithms},\ }\textbf {\bibinfo {volume} {6}},\ \bibinfo {pages} {161}
  (\bibinfo {year} {1995})}\BibitemShut {NoStop}%
\bibitem [{\citenamefont {Albert}\ \emph {et~al.}(2000)\citenamefont {Albert},
  \citenamefont {Jeong},\ and\ \citenamefont
  {Barab\'{a}si}}]{albert:2000_error}%
  \BibitemOpen
  \bibfield  {author} {\bibinfo {author} {\bibfnamefont {R.}~\bibnamefont
  {Albert}}, \bibinfo {author} {\bibfnamefont {H.}~\bibnamefont {Jeong}}, \
  and\ \bibinfo {author} {\bibfnamefont {A.-L.}\ \bibnamefont {Barab\'{a}si}},\
  }\href@noop {} {\bibfield  {journal} {\bibinfo  {journal} {Nature (London)},\
  }\textbf {\bibinfo {volume} {406}},\ \bibinfo {pages} {378} (\bibinfo {year}
  {2000})}\BibitemShut {NoStop}%
\bibitem [{\citenamefont {Cohen}\ \emph {et~al.}(2000)\citenamefont {Cohen},
  \citenamefont {Erez}, \citenamefont {ben Avraham},\ and\ \citenamefont
  {Havlin}}]{Cohen:2000vq}%
  \BibitemOpen
  \bibfield  {author} {\bibinfo {author} {\bibfnamefont {R.}~\bibnamefont
  {Cohen}}, \bibinfo {author} {\bibfnamefont {K.}~\bibnamefont {Erez}},
  \bibinfo {author} {\bibfnamefont {D.}~\bibnamefont {ben Avraham}}, \ and\
  \bibinfo {author} {\bibfnamefont {S.}~\bibnamefont {Havlin}},\ }\href@noop {}
  {\bibfield  {journal} {\bibinfo  {journal} {Phys. Rev. Lett.},\ }\textbf
  {\bibinfo {volume} {85}},\ \bibinfo {pages} {4626} (\bibinfo {year}
  {2000})}\BibitemShut {NoStop}%
\bibitem [{\citenamefont {Moore}\ and\ \citenamefont
  {Newman}(2000)}]{Moore:2000ve}%
  \BibitemOpen
  \bibfield  {author} {\bibinfo {author} {\bibfnamefont {C.}~\bibnamefont
  {Moore}}\ and\ \bibinfo {author} {\bibfnamefont {M.~E.~J.}\ \bibnamefont
  {Newman}},\ }\href@noop {} {\bibfield  {journal} {\bibinfo  {journal} {Phys.
  Rev. E},\ }\textbf {\bibinfo {volume} {62}},\ \bibinfo {pages} {7059}
  (\bibinfo {year} {2000})}\BibitemShut {NoStop}%
\bibitem [{\citenamefont {Callaway}\ \emph {et~al.}(2000)\citenamefont
  {Callaway}, \citenamefont {Newman}, \citenamefont {Strogatz},\ and\
  \citenamefont {Watts}}]{Callaway:2000vd}%
  \BibitemOpen
  \bibfield  {author} {\bibinfo {author} {\bibfnamefont {D.~S.}\ \bibnamefont
  {Callaway}}, \bibinfo {author} {\bibfnamefont {M.~E.~J.}\ \bibnamefont
  {Newman}}, \bibinfo {author} {\bibfnamefont {S.~H.}\ \bibnamefont
  {Strogatz}}, \ and\ \bibinfo {author} {\bibfnamefont {D.~J.}\ \bibnamefont
  {Watts}},\ }\href@noop {} {\bibfield  {journal} {\bibinfo  {journal} {Phys.
  Rev. Lett.},\ }\textbf {\bibinfo {volume} {85}},\ \bibinfo {pages} {5468}
  (\bibinfo {year} {2000})}\BibitemShut {NoStop}%
\bibitem [{\citenamefont {Cohen}\ \emph {et~al.}(2001)\citenamefont {Cohen},
  \citenamefont {Erez}, \citenamefont {ben Avraham},\ and\ \citenamefont
  {Havlin}}]{Cohen:2001hf}%
  \BibitemOpen
  \bibfield  {author} {\bibinfo {author} {\bibfnamefont {R.}~\bibnamefont
  {Cohen}}, \bibinfo {author} {\bibfnamefont {K.}~\bibnamefont {Erez}},
  \bibinfo {author} {\bibfnamefont {D.}~\bibnamefont {ben Avraham}}, \ and\
  \bibinfo {author} {\bibfnamefont {S.}~\bibnamefont {Havlin}},\ }\href@noop {}
  {\bibfield  {journal} {\bibinfo  {journal} {Phys. Rev. Lett.},\ }\textbf
  {\bibinfo {volume} {86}},\ \bibinfo {pages} {3682} (\bibinfo {year}
  {2001})}\BibitemShut {NoStop}%
\bibitem [{\citenamefont {Cohen}\ \emph {et~al.}(2002)\citenamefont {Cohen},
  \citenamefont {ben Avraham},\ and\ \citenamefont {Havlin}}]{Cohen:2002br}%
  \BibitemOpen
  \bibfield  {author} {\bibinfo {author} {\bibfnamefont {R.}~\bibnamefont
  {Cohen}}, \bibinfo {author} {\bibfnamefont {D.}~\bibnamefont {ben Avraham}},
  \ and\ \bibinfo {author} {\bibfnamefont {S.}~\bibnamefont {Havlin}},\
  }\href@noop {} {\bibfield  {journal} {\bibinfo  {journal} {Phys. Rev. E},\
  }\textbf {\bibinfo {volume} {66}},\ \bibinfo {pages} {036113} (\bibinfo
  {year} {2002})}\BibitemShut {NoStop}%
\bibitem [{\citenamefont {Chung}\ and\ \citenamefont
  {Lu}(2002)}]{Chung:2002wu}%
  \BibitemOpen
  \bibfield  {author} {\bibinfo {author} {\bibfnamefont {F.}~\bibnamefont
  {Chung}}\ and\ \bibinfo {author} {\bibfnamefont {L.}~\bibnamefont {Lu}},\
  }\href@noop {} {\bibfield  {journal} {\bibinfo  {journal} {Annals of
  Combinatorics},\ \bibinfo {pages} {125}} (\bibinfo {year}
  {2002})}\BibitemShut {NoStop}%
\bibitem [{\citenamefont {Schwartz}\ \emph {et~al.}(2002)\citenamefont
  {Schwartz}, \citenamefont {Cohen}, \citenamefont {ben Avraham}, \citenamefont
  {Barab\'{a}si},\ and\ \citenamefont {Havlin}}]{Schwartz:2002dn}%
  \BibitemOpen
  \bibfield  {author} {\bibinfo {author} {\bibfnamefont {N.}~\bibnamefont
  {Schwartz}}, \bibinfo {author} {\bibfnamefont {R.}~\bibnamefont {Cohen}},
  \bibinfo {author} {\bibfnamefont {D.}~\bibnamefont {ben Avraham}}, \bibinfo
  {author} {\bibfnamefont {A.-L.}\ \bibnamefont {Barab\'{a}si}}, \ and\
  \bibinfo {author} {\bibfnamefont {S.}~\bibnamefont {Havlin}},\ }\href@noop {}
  {\bibfield  {journal} {\bibinfo  {journal} {Phys. Rev. E},\ }\textbf
  {\bibinfo {volume} {66}},\ \bibinfo {pages} {015104} (\bibinfo {year}
  {2002})}\BibitemShut {NoStop}%
\bibitem [{\citenamefont {Gallos}\ \emph {et~al.}(2005)\citenamefont {Gallos},
  \citenamefont {Cohen}, \citenamefont {Argyrakis}, \citenamefont {Bunde},\
  and\ \citenamefont {Havlin}}]{Gallos:2005gs}%
  \BibitemOpen
  \bibfield  {author} {\bibinfo {author} {\bibfnamefont {L.~K.}\ \bibnamefont
  {Gallos}}, \bibinfo {author} {\bibfnamefont {R.}~\bibnamefont {Cohen}},
  \bibinfo {author} {\bibfnamefont {P.}~\bibnamefont {Argyrakis}}, \bibinfo
  {author} {\bibfnamefont {A.}~\bibnamefont {Bunde}}, \ and\ \bibinfo {author}
  {\bibfnamefont {S.}~\bibnamefont {Havlin}},\ }\href@noop {} {\bibfield
  {journal} {\bibinfo  {journal} {Phys. Rev. Lett.},\ }\textbf {\bibinfo
  {volume} {94}},\ \bibinfo {pages} {188701} (\bibinfo {year}
  {2005})}\BibitemShut {NoStop}%
\bibitem [{\citenamefont {Shargel}\ \emph {et~al.}(2003)\citenamefont
  {Shargel}, \citenamefont {Sayama}, \citenamefont {Epstein},\ and\
  \citenamefont {Bar-Yam}}]{Shargel:2003cu}%
  \BibitemOpen
  \bibfield  {author} {\bibinfo {author} {\bibfnamefont {B.}~\bibnamefont
  {Shargel}}, \bibinfo {author} {\bibfnamefont {H.}~\bibnamefont {Sayama}},
  \bibinfo {author} {\bibfnamefont {I.~R.}\ \bibnamefont {Epstein}}, \ and\
  \bibinfo {author} {\bibfnamefont {Y.}~\bibnamefont {Bar-Yam}},\ }\href@noop
  {} {\bibfield  {journal} {\bibinfo  {journal} {Phys. Rev. Lett.},\ }\textbf
  {\bibinfo {volume} {90}},\ \bibinfo {pages} {068701} (\bibinfo {year}
  {2003})}\BibitemShut {NoStop}%
\bibitem [{\citenamefont {V{\'a}zquez}\ and\ \citenamefont
  {Moreno}(2003)}]{Vazquez:2003fq}%
  \BibitemOpen
  \bibfield  {author} {\bibinfo {author} {\bibfnamefont {A.}~\bibnamefont
  {V{\'a}zquez}}\ and\ \bibinfo {author} {\bibfnamefont {Y.}~\bibnamefont
  {Moreno}},\ }\href@noop {} {\bibfield  {journal} {\bibinfo  {journal} {Phys.
  Rev. E},\ }\textbf {\bibinfo {volume} {67}},\ \bibinfo {pages} {015101}
  (\bibinfo {year} {2003})}\BibitemShut {NoStop}%
\bibitem [{\citenamefont {Paul}\ \emph {et~al.}(2004)\citenamefont {Paul},
  \citenamefont {Tanizawa}, \citenamefont {Havlin},\ and\ \citenamefont
  {Stanley}}]{Paul:2004br}%
  \BibitemOpen
  \bibfield  {author} {\bibinfo {author} {\bibfnamefont {G.}~\bibnamefont
  {Paul}}, \bibinfo {author} {\bibfnamefont {T.}~\bibnamefont {Tanizawa}},
  \bibinfo {author} {\bibfnamefont {S.}~\bibnamefont {Havlin}}, \ and\ \bibinfo
  {author} {\bibfnamefont {H.~E.}\ \bibnamefont {Stanley}},\ }\href@noop {}
  {\bibfield  {journal} {\bibinfo  {journal} {Eur. Phys. J. B},\ }\textbf
  {\bibinfo {volume} {38}},\ \bibinfo {pages} {187} (\bibinfo {year}
  {2004})}\BibitemShut {NoStop}%
\bibitem [{\citenamefont {Tanizawa}\ \emph {et~al.}(2005)\citenamefont
  {Tanizawa}, \citenamefont {Paul}, \citenamefont {Cohen}, \citenamefont
  {Havlin},\ and\ \citenamefont {Stanley}}]{Tanizawa:2005fd}%
  \BibitemOpen
  \bibfield  {author} {\bibinfo {author} {\bibfnamefont {T.}~\bibnamefont
  {Tanizawa}}, \bibinfo {author} {\bibfnamefont {G.}~\bibnamefont {Paul}},
  \bibinfo {author} {\bibfnamefont {R.}~\bibnamefont {Cohen}}, \bibinfo
  {author} {\bibfnamefont {S.}~\bibnamefont {Havlin}}, \ and\ \bibinfo {author}
  {\bibfnamefont {H.~E.}\ \bibnamefont {Stanley}},\ }\href@noop {} {\bibfield
  {journal} {\bibinfo  {journal} {Phys. Rev. E},\ }\textbf {\bibinfo {volume}
  {71}},\ \bibinfo {pages} {047101} (\bibinfo {year} {2005})}\BibitemShut
  {NoStop}%
\bibitem [{\citenamefont {Tanizawa}\ \emph {et~al.}(2006)\citenamefont
  {Tanizawa}, \citenamefont {Paul}, \citenamefont {Havlin},\ and\ \citenamefont
  {Stanley}}]{Tanizawa:2006dn}%
  \BibitemOpen
  \bibfield  {author} {\bibinfo {author} {\bibfnamefont {T.}~\bibnamefont
  {Tanizawa}}, \bibinfo {author} {\bibfnamefont {G.}~\bibnamefont {Paul}},
  \bibinfo {author} {\bibfnamefont {S.}~\bibnamefont {Havlin}}, \ and\ \bibinfo
  {author} {\bibfnamefont {H.~E.}\ \bibnamefont {Stanley}},\ }\href@noop {}
  {\bibfield  {journal} {\bibinfo  {journal} {Phys. Rev. E},\ }\textbf
  {\bibinfo {volume} {74}},\ \bibinfo {pages} {016125} (\bibinfo {year}
  {2006})}\BibitemShut {NoStop}%
\bibitem [{\citenamefont {Paul}\ \emph {et~al.}(2006)\citenamefont {Paul},
  \citenamefont {Sreenivasan}, \citenamefont {Havlin},\ and\ \citenamefont
  {Stanley}}]{Paul:2006cd}%
  \BibitemOpen
  \bibfield  {author} {\bibinfo {author} {\bibfnamefont {G.}~\bibnamefont
  {Paul}}, \bibinfo {author} {\bibfnamefont {S.}~\bibnamefont {Sreenivasan}},
  \bibinfo {author} {\bibfnamefont {S.}~\bibnamefont {Havlin}}, \ and\ \bibinfo
  {author} {\bibfnamefont {H.~E.}\ \bibnamefont {Stanley}},\ }\href@noop {}
  {\bibfield  {journal} {\bibinfo  {journal} {Physica A},\ }\textbf {\bibinfo
  {volume} {370}},\ \bibinfo {pages} {854} (\bibinfo {year}
  {2006})}\BibitemShut {NoStop}%
\bibitem [{\citenamefont {Donetti}\ \emph {et~al.}(2006)\citenamefont
  {Donetti}, \citenamefont {Neri},\ and\ \citenamefont
  {Mu{\~n}oz}}]{Donetti:2006ck}%
  \BibitemOpen
  \bibfield  {author} {\bibinfo {author} {\bibfnamefont {L.}~\bibnamefont
  {Donetti}}, \bibinfo {author} {\bibfnamefont {F.}~\bibnamefont {Neri}}, \
  and\ \bibinfo {author} {\bibfnamefont {M.}~\bibnamefont {Mu{\~n}oz}},\
  }\href@noop {} {\bibfield  {journal} {\bibinfo  {journal} {Journal of
  Statistical Mechanics: Theory and Experiment},\ \bibinfo {pages} {P08007}}
  (\bibinfo {year} {2006})}\BibitemShut {NoStop}%
\bibitem [{\citenamefont {Paul}\ \emph {et~al.}(2007)\citenamefont {Paul},
  \citenamefont {Cohen}, \citenamefont {Sreenivasan}, \citenamefont {Havlin},\
  and\ \citenamefont {Stanley}}]{Paul:2007br}%
  \BibitemOpen
  \bibfield  {author} {\bibinfo {author} {\bibfnamefont {G.}~\bibnamefont
  {Paul}}, \bibinfo {author} {\bibfnamefont {R.}~\bibnamefont {Cohen}},
  \bibinfo {author} {\bibfnamefont {S.}~\bibnamefont {Sreenivasan}}, \bibinfo
  {author} {\bibfnamefont {S.}~\bibnamefont {Havlin}}, \ and\ \bibinfo {author}
  {\bibfnamefont {H.~E.}\ \bibnamefont {Stanley}},\ }\href@noop {} {\bibfield
  {journal} {\bibinfo  {journal} {Phys. Rev. Lett.},\ }\textbf {\bibinfo
  {volume} {99}},\ \bibinfo {pages} {115701} (\bibinfo {year}
  {2007})}\BibitemShut {NoStop}%
\bibitem [{\citenamefont {Buldyrev}\ \emph {et~al.}(2010)\citenamefont
  {Buldyrev}, \citenamefont {Parshani}, \citenamefont {Paul}, \citenamefont
  {Stanley},\ and\ \citenamefont {Havlin}}]{buldyrev:2010_catas}%
  \BibitemOpen
  \bibfield  {author} {\bibinfo {author} {\bibfnamefont {S.~V.}\ \bibnamefont
  {Buldyrev}}, \bibinfo {author} {\bibfnamefont {R.}~\bibnamefont {Parshani}},
  \bibinfo {author} {\bibfnamefont {G.}~\bibnamefont {Paul}}, \bibinfo {author}
  {\bibfnamefont {H.~E.}\ \bibnamefont {Stanley}}, \ and\ \bibinfo {author}
  {\bibfnamefont {S.}~\bibnamefont {Havlin}},\ }\href@noop {} {\bibfield
  {journal} {\bibinfo  {journal} {Nature (London)},\ }\textbf {\bibinfo
  {volume} {464}},\ \bibinfo {pages} {1025} (\bibinfo {year}
  {2010})}\BibitemShut {NoStop}%
\bibitem [{\citenamefont {Huang}\ \emph {et~al.}(2011)\citenamefont {Huang},
  \citenamefont {Gao}, \citenamefont {Buldyrev}, \citenamefont {Havlin},\ and\
  \citenamefont {Stanley}}]{Huang:2011cg}%
  \BibitemOpen
  \bibfield  {author} {\bibinfo {author} {\bibfnamefont {X.}~\bibnamefont
  {Huang}}, \bibinfo {author} {\bibfnamefont {J.}~\bibnamefont {Gao}}, \bibinfo
  {author} {\bibfnamefont {S.}~\bibnamefont {Buldyrev}}, \bibinfo {author}
  {\bibfnamefont {S.}~\bibnamefont {Havlin}}, \ and\ \bibinfo {author}
  {\bibfnamefont {H.~E.}\ \bibnamefont {Stanley}},\ }\href@noop {} {\bibfield
  {journal} {\bibinfo  {journal} {Phys. Rev. E},\ }\textbf {\bibinfo {volume}
  {83}} (\bibinfo {year} {2011})}\BibitemShut {NoStop}%
\bibitem [{\citenamefont {Serrano}\ \emph {et~al.}(2006)\citenamefont
  {Serrano}, \citenamefont {Bogu{\~n}{\'a}},\ and\ \citenamefont
  {Pastor-Satorras}}]{Serrano:2006ik}%
  \BibitemOpen
  \bibfield  {author} {\bibinfo {author} {\bibfnamefont {M.~A.}\ \bibnamefont
  {Serrano}}, \bibinfo {author} {\bibfnamefont {M.}~\bibnamefont
  {Bogu{\~n}{\'a}}}, \ and\ \bibinfo {author} {\bibfnamefont {R.}~\bibnamefont
  {Pastor-Satorras}},\ }\href@noop {} {\bibfield  {journal} {\bibinfo
  {journal} {Phys. Rev. E},\ }\textbf {\bibinfo {volume} {74}},\ \bibinfo
  {pages} {055101} (\bibinfo {year} {2006})}\BibitemShut {NoStop}%
\bibitem [{\citenamefont {Goltsev}\ \emph {et~al.}(2008)\citenamefont
  {Goltsev}, \citenamefont {Dorogovtsev},\ and\ \citenamefont
  {Mendes}}]{Goltsev:2008bf}%
  \BibitemOpen
  \bibfield  {author} {\bibinfo {author} {\bibfnamefont {A.~V.}\ \bibnamefont
  {Goltsev}}, \bibinfo {author} {\bibfnamefont {S.~N.}\ \bibnamefont
  {Dorogovtsev}}, \ and\ \bibinfo {author} {\bibfnamefont {J.~F.~F.}\
  \bibnamefont {Mendes}},\ }\href@noop {} {\bibfield  {journal} {\bibinfo
  {journal} {Phys. Rev. E},\ }\textbf {\bibinfo {volume} {78}},\ \bibinfo
  {pages} {051105} (\bibinfo {year} {2008})}\BibitemShut {NoStop}%
\bibitem [{\citenamefont {Shiraki}\ and\ \citenamefont
  {Kabashima}(2010)}]{Shiraki:2010is}%
  \BibitemOpen
  \bibfield  {author} {\bibinfo {author} {\bibfnamefont {Y.}~\bibnamefont
  {Shiraki}}\ and\ \bibinfo {author} {\bibfnamefont {Y.}~\bibnamefont
  {Kabashima}},\ }\href@noop {} {\bibfield  {journal} {\bibinfo  {journal}
  {Phys. Rev. E},\ }\textbf {\bibinfo {volume} {82}},\ \bibinfo {pages}
  {036101} (\bibinfo {year} {2010})}\BibitemShut {NoStop}%
\bibitem [{\citenamefont {Ostilli}\ \emph {et~al.}(2011)\citenamefont
  {Ostilli}, \citenamefont {Ferreira},\ and\ \citenamefont
  {Mendes}}]{Ostilli:2011hb}%
  \BibitemOpen
  \bibfield  {author} {\bibinfo {author} {\bibfnamefont {M.}~\bibnamefont
  {Ostilli}}, \bibinfo {author} {\bibfnamefont {A.}~\bibnamefont {Ferreira}}, \
  and\ \bibinfo {author} {\bibfnamefont {J.}~\bibnamefont {Mendes}},\
  }\href@noop {} {\bibfield  {journal} {\bibinfo  {journal} {Phys. Rev. E},\
  }\textbf {\bibinfo {volume} {83}} (\bibinfo {year} {2011})}\BibitemShut
  {NoStop}%
\bibitem [{\citenamefont {Tanizawa}\ \emph {et~al.}(2012)\citenamefont
  {Tanizawa}, \citenamefont {Havlin},\ and\ \citenamefont
  {Stanley}}]{Tanizawa:2012hh}%
  \BibitemOpen
  \bibfield  {author} {\bibinfo {author} {\bibfnamefont {T.}~\bibnamefont
  {Tanizawa}}, \bibinfo {author} {\bibfnamefont {S.}~\bibnamefont {Havlin}}, \
  and\ \bibinfo {author} {\bibfnamefont {H.~E.}\ \bibnamefont {Stanley}},\
  }\href@noop {} {\bibfield  {journal} {\bibinfo  {journal} {Phys. Rev. E},\
  }\textbf {\bibinfo {volume} {85}},\ \bibinfo {pages} {046109} (\bibinfo
  {year} {2012})}\BibitemShut {NoStop}%
\bibitem [{\citenamefont {Newman}(2002)}]{Newman:2002jj}%
  \BibitemOpen
  \bibfield  {author} {\bibinfo {author} {\bibfnamefont {M.~E.~J.}\
  \bibnamefont {Newman}},\ }\href@noop {} {\bibfield  {journal} {\bibinfo
  {journal} {Phys. Rev. Lett.},\ }\textbf {\bibinfo {volume} {89}},\ \bibinfo
  {pages} {208701} (\bibinfo {year} {2002})}\BibitemShut {NoStop}%
\bibitem [{\citenamefont {Schneider}\ \emph {et~al.}(2011)\citenamefont
  {Schneider}, \citenamefont {Moreira}, \citenamefont {Andrade~Jr},
  \citenamefont {Havlin},\ and\ \citenamefont {Herrmann}}]{Schneider:2011ip}%
  \BibitemOpen
  \bibfield  {author} {\bibinfo {author} {\bibfnamefont {C.~M.}\ \bibnamefont
  {Schneider}}, \bibinfo {author} {\bibfnamefont {A.~A.}\ \bibnamefont
  {Moreira}}, \bibinfo {author} {\bibfnamefont {J.~S.}\ \bibnamefont
  {Andrade~Jr}}, \bibinfo {author} {\bibfnamefont {S.}~\bibnamefont {Havlin}},
  \ and\ \bibinfo {author} {\bibfnamefont {H.~J.}\ \bibnamefont {Herrmann}},\
  }\href@noop {} {\bibfield  {journal} {\bibinfo  {journal} {PNAS},\ }\textbf
  {\bibinfo {volume} {108}},\ \bibinfo {pages} {3838} (\bibinfo {year}
  {2011})}\BibitemShut {NoStop}%
\bibitem [{\citenamefont {Herrmann}\ \emph {et~al.}(2011)\citenamefont
  {Herrmann}, \citenamefont {Schneider}, \citenamefont {Moreira}, \citenamefont
  {Andrade~Jr},\ and\ \citenamefont {Havlin}}]{Herrmann:2011hd}%
  \BibitemOpen
  \bibfield  {author} {\bibinfo {author} {\bibfnamefont {H.~J.}\ \bibnamefont
  {Herrmann}}, \bibinfo {author} {\bibfnamefont {C.~M.}\ \bibnamefont
  {Schneider}}, \bibinfo {author} {\bibfnamefont {A.~A.}\ \bibnamefont
  {Moreira}}, \bibinfo {author} {\bibfnamefont {J.~S.}\ \bibnamefont
  {Andrade~Jr}}, \ and\ \bibinfo {author} {\bibfnamefont {S.}~\bibnamefont
  {Havlin}},\ }\href@noop {} {\bibfield  {journal} {\bibinfo  {journal}
  {Journal of Statistical Mechanics: Theory and Experiment},\ }\textbf
  {\bibinfo {volume} {2011}},\ \bibinfo {pages} {P01027} (\bibinfo {year}
  {2011})}\BibitemShut {NoStop}%
\bibitem [{\citenamefont {Gallos}\ \emph {et~al.}(2008)\citenamefont {Gallos},
  \citenamefont {Song},\ and\ \citenamefont {Makse}}]{Gallos:2008hd}%
  \BibitemOpen
  \bibfield  {author} {\bibinfo {author} {\bibfnamefont {L.~K.}\ \bibnamefont
  {Gallos}}, \bibinfo {author} {\bibfnamefont {C.}~\bibnamefont {Song}}, \ and\
  \bibinfo {author} {\bibfnamefont {H.~A.}\ \bibnamefont {Makse}},\ }\href@noop
  {} {\bibfield  {journal} {\bibinfo  {journal} {Phys. Rev. Lett.}} (\bibinfo
  {year} {2008})}\BibitemShut {NoStop}%
\bibitem [{Note1()}]{Note1}%
  \BibitemOpen
  \bibinfo {note} {Gallos et al.\ took a similar form for the joint degree
  probability \cite {Gallos:2008hd}. It should be noted, however, that they
  gave the definition for $q \le k$ and the definition of $\epsilon $ is
  different.}\BibitemShut {Stop}%
\bibitem [{Note2()}]{Note2}%
  \BibitemOpen
  \bibinfo {note} {This is the same equation that Gallos et al.\ examined in a
  different way \cite {Gallos:2008hd}.}\BibitemShut {Stop}%
\bibitem [{Note3()}]{Note3}%
  \BibitemOpen
  \bibinfo {note} {It is well known that $\lambda $ takes a value between 2 and
  3 for many scale-free networks in the real world. Since the main focus of
  this paper is to investigate the effects of disassortativity, we take
  $\lambda = 1.0$ in this paper to make the number of hubs larger and the
  effects of disassortativity much more prominent. We also calculated for the
  case of scale-free networks with $m = 2$, $K = 20$, and $\lambda = 2.2$ and
  obtained similar results.}\BibitemShut {Stop}%
\bibitem [{Note4()}]{Note4}%
  \BibitemOpen
  \bibinfo {note} {This measure is the same as those used in \cite
  {Schneider:2011ip,Herrmann:2011hd,Tanizawa:2012hh}}\BibitemShut {NoStop}%
\bibitem [{Note5()}]{Note5}%
  \BibitemOpen
  \bibinfo {note} {The density, $\rho _k(t)$, in the diffusion equation, Eq.\
  (\ref {eq:diffusion}), is the density of ``particles'' on each node. Since
  the number of nodes of $m$ degree is much larger than the number of nodes of
  $K$ degree in scale-free networks, the final values of $\rho _m$ plotted in
  Fig.\ \ref {fig:diffusion Km} (a) are much smaller than the final values of
  $\rho _K$ plotted in Fig.\ \ref {fig:diffusion mK} (b).}\BibitemShut {Stop}%
\end{thebibliography}

%

\end{document}